\documentclass[a4paper]{article}
\usepackage[english]{babel}
\usepackage[T2A]{fontenc}
\usepackage[utf8]{inputenc}
\usepackage{amsmath}
\usepackage{amsthm}
\usepackage{amssymb}
\usepackage{textcomp}
\usepackage{graphicx}
\usepackage{array}
\usepackage{hyperref}
\usepackage{tikz}
\usepackage{float}
\usepackage{authblk}
\usetikzlibrary{decorations.markings}
\usetikzlibrary{shapes}

\newcommand{\be}{\begin{eqnarray}}
\newcommand{\ee}{\end{eqnarray}}

\newcommand{\eqlb}[2]{\begin{equation} \label{#1} #2 \end{equation}}
\newcommand{\eq}[1]{\begin{equation} #1 \end{equation}}
\newcommand{\eqn}[1]{\begin{equation*} #1 \end{equation*}}

\newcommand{\eqs}[1]{$#1$}
\newcommand{\brc}[1]{\left(#1\right)}
\newcommand{\bsq}[1]{\left[#1\right]}

\newcommand{\wt}[1]{\widetilde{#1}}

\newcommand{\bld}{\boldsymbol}

\newcommand{\rme}{\textrm{e}}
\newcommand{\rmd}{\textrm{d}}
\newcommand{\Tr}{\textrm{Tr}}

\numberwithin{equation}{section}

\hoffset= - 1.0in
\begin{document}

\title{Spontaneous symmetry breaking in pure $2$D Yang-Mills theory}
\author[a,b]{G. Aminov\thanks{gleb.aminov@stonybrook.edu}}
\affil[a]{\small Department of Physics and Astronomy, Stony Brook University, Stony Brook, NY 11794, USA}
\affil[b]{\small ITEP NRC KI, Moscow 117218, Russia}

\date{ }

\maketitle

\abstract
We consider purely topological $2$d Yang-Mills theory on a torus with the second Stiefel–Whitney class added to the Lagrangian in the form of a \eqs{\theta}-term. It will be shown, that at \eqs{\theta=\pi} there exists a class of \eqs{SU(2 N)/\mathbb{Z}_2} (\eqs{N>1}) gauge theories with a two-fold degenerate vacuum, which spontaneously breaks the time reversal and charge conjugation symmetries. The corresponding order parameter is given by the generator \eqs{\mathcal{O}} of the $\mathbb{Z}_N$ one-form symmetry.

\section{Introduction}
The possibility of having a number of degenerate vacua called $\theta$-vacua in two dimensional gauge theories was
studied in the $70$'s by a number of authors \cite{LS'71,Hooft'74,Coleman'76,CDG'76,Polyakov'77,Witten'79}. Both abelian and non-abelian theories were considered and the existence of the multiple vacua was shown to be independent of the spontaneous symmetry breaking of the gauge symmetry. Instead, the presence of some matter fields, either fermionic or scalar, was required.

In this note we consider purely topological $2$d Yang-Mills theory on a torus with the second Stiefel–Whitney class added to the Lagrangian in the form of a \eqs{\theta}-term. It will be shown, that at \eqs{\theta=\pi} there exists a class of \eqs{SU(2 N)/\mathbb{Z}_2} (\eqs{N>1}) gauge theories with a two-fold degenerate vacuum. These two vacuum states are related by the time reversal or the charge conjugation and thus indicate the spontaneous symmetry breaking. The corresponding order parameter is given by the generator \eqs{\mathcal{O}} of the $\mathbb{Z}_N$ one-form symmetry with the following action of the charge conjugation on it:
\eq{\textbf{C}\,\mathcal{O}\,\textbf{C}^{-1}=\mathcal{O}^{-1}.}
The motivation to consider such theories comes from the recent developments in generalized global symmetries and 't~Hooft anomalies \cite{KS'14,GKSW'15,Hooft'80,Redlich1'84,Redlich2'84,KT1'14,KT2'14,SW'16,BHS'17,GKKS'17,TMS'17,KSTZ'18, KSU'18,TKMS'18,MTU'19,CFLS'19,CJTU'19,CO'19,KMS'19,WWZ'19,WWZ'20}. In particular, authors of \cite{GKKS'17} considered \eqs{SU(N)} gauge theory in $4$ dimensions and showed that at \eqs{\theta=\pi} there is the discrete `t~Hooft anomaly involving time reversal and the center symmetry. As a consequence of this anomaly, the vacuum at \eqs{\theta=\pi} cannot be a trivial non-degenerate gapped state. Another example of `t~Hooft anomaly constraining the vacuum of the theory is related to the $2$d \eqs{\mathbb{CP}^{n-1}} model \cite{BHS'17}, where for \eqs{n>2} the mixed anomaly between time reversal symmetry and the global \eqs{PSU\brc{n}} symmetry at \eqs{\theta=\pi} leads to the spontaneous breaking of time reversal symmetry with a two-fold degeneracy of the vacuum \cite{Affleck'91}. The list of the examples could be made longer, but we will conclude by mentioning the works \cite{KT1'14,KT2'14}, where the `t~Hooft anomalies for discrete global symmetries in
bosonic theories were studied in $2$, $3$ and $4$ dimensions. Although in this note we are not going to discuss possible relation of the spontaneous symmetry breaking to the anomaly, but one could hypothesize the existence of the mixed anomaly between $\brc{-1}$-form symmetry and the charge conjugation in the theories under consideration \footnote{This possibility was pointed out to the author by Z.~Komargodski}.

While the note was in preparation, we became aware of the paper by D.~Kapec, R.~Mahajan and D.~Stanford \cite{KMS'19},
which has partial overlap with our results for the partition functions of \eqs{PSU\brc{N}} gauge theories. In \cite{KMS'19} the higher genus partition functions were computed and utilized in the context of random matrix ensembles. As we will show in the main text of the note, there is no spontaneous symmetry breaking for the case of \eqs{PSU\brc{N}} gauge theories and thus the main results of our study are not covered in \cite{KMS'19}. Also, the paper by E.~Sharpe \cite{Sharpe'19} discussing $1$-form symmetries in the various $2$d theories appeared soon after the first draft of this note. This paper studies the connection with the cluster decomposition and is based on a number of previous results (to name a few \cite{PS'05,HHPSA'06,Sharpe'14}).

The note is organized as follows. In section \ref{sec:review} we review the Hamiltonian approach for computing the partition functions of the pure gauge theories in two dimensions. This method originates from the work of A.~Migdal \cite{Migdal'75} and was extensively developed in the $80$'s and $90$'s alongside other approaches for studying the  $2$d Yang-Mills theories \cite{KK'80,Kazakov'81,AB'83,BZ'85,Fine'90,Fine'91,Witten'91,GT'93,Kostov'94,MP'94,CMR'97,CMR'95, Kazakov'94}. We would also like to mention the path integral approach by M.~Blau and G.~Thompson \cite{BT'91,BT'92,BT'93}, which leads to the same results, but requires more involved mathematical structures. In section \ref{sec:PSU(2)} we use G.~'t~Hooft's twisted boundary conditions \cite{Hooft'79} to compute the partition function of the \eqs{SU(2)/\mathbb{Z}_2} gauge theory. This computation is equivalent to the approach used by E.~Witten \cite{Witten'91} to compute the \eqs{SO\brc{3}} partition function starting from the \eqs{SU(2)} gauge theory. We conclude section \ref{sec:PSU(2)} by introducing the \eqs{\theta}-term to the Lagrangian and computing the partition function at \eqs{\theta = \pi}, which repeats one of the results of \cite{Alekseev'15} and \cite{Sharpe'14}. In section \ref{sec:PSU(N)} we extend all the previous arguments to the case of \eqs{SU(N)/\mathbb{Z}_N} theory. However, since there is no spontaneous symmetry breaking in \eqs{PSU\brc{N}} theory for any $N$, we switch in section \ref{sec:2vacua} to the more general case of \eqs{SU\brc{N}/\Gamma}, where \eqs{\Gamma} is the subgroup of the center of \eqs{SU\brc{N}}. Indeed, we find out that there exists a class of \eqs{SU(2 N)/\mathbb{Z}_2}, \eqs{N>1} theories with two vacuum states given by the fundamental and antifundamental representations of \eqs{SU(2 N)}. Additionally, we argue that there exists a broader class of \eqs{SU(2N\,m)/\mathbb{Z}_{2m}} theories with degenerate vacuum. Finally, in section \ref{sec:SSB} we relate the two-fold degeneracy of the vacuum to the spontaneous breaking of \eqs{\textbf{C}} and \eqs{\textbf{T}} symmetries.

\section{Review: \eqs{SU(2)} gauge theory}
\label{sec:review}
To derive the answer for the partition function on the torus we consider the canonical quantization
of the theory on a cylinder (See Figure 1).
\begin{figure}[H]
  \begin{center}
  \includegraphics[width=3.5cm]{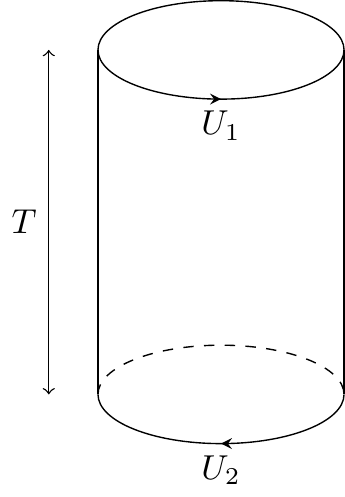}
  \caption{Holonomies on the cylinder}
  \end{center}
\end{figure}
The corresponding propagator \cite{Migdal'75,Witten'91,CMR'95} is given by
\eqlb{eq:propcyl}{Z\brc{a,U_1,U_2}=\sum_{R} \chi_{R}\brc{U_1}\, \chi_{R}\brc{U_2} \rme^{-a C_2\brc{R}},}
where \eqs{a=e^2 L T/2} is proportional to the surface area of the cylinder. The final answer for the partition function on the torus comes from gluing together the opposite sides of the cylinder:
\eq{Z=\int \rmd U_1\, Z\brc{a,U_1,U_1^{-1}} = \sum_{R}  \rme^{-a C_2\brc{R}}
\int \rmd U_1\,\chi_{R}\brc{U_1} \chi_{R}\brc{U_1^{-1}}.}
Using the identity
\eq{\int\rmd U \, \chi_R\brc{V U} \chi_R\brc{U^{-1} W}=\frac{\chi_R \brc{V W}}{\textrm{dim} \,R},}
we get
\eqlb{eq:Z_2D}{Z=\sum_{R} \rme^{-a C_2\brc{R}}.}
For \eqs{SU\brc{2}} we have $C_2\brc{R}=j\brc{j+1}$ with half-integer $j$ and hence
\eq{Z=\sum_{m=0}^{\infty}\rme^{-a\, m\brc{m+2}/4}.}

\section{\eqs{SU(2)/\mathbb{Z}_2} gauge theory}
\label{sec:PSU(2)}
Now, we consider the cylinder as a rectangular plaquette with one pair of opposite sides being
glued together. According to \cite{Hooft'79} we can introduce the following boundary conditions for the vector potential \eqs{A_{\mu}\brc{x,t}}:
\eq{\left\{
\begin{array}{l}
A_{\mu}\brc{L,t}=\wt\Omega_1\brc{t} A_{\mu}\brc{0,t},\\
A_{\mu}\brc{x,T}=\wt\Omega_2\brc{x} A_{\mu}\brc{x,0},
\end{array}\right.
}
with the notation \eqs{\Omega A_{\mu}=\Omega A_{\mu} \Omega^{-1} +\frac{\imath}{g}\Omega\partial_{\mu} \Omega^{-1}}.
However, since we are using the \eqs{A_0=0} gauge, we are left with time-independent gauge transformations:
\eq{\wt\Omega_1\brc{t}=\wt\Omega_1\brc{0}.}
Now, making a constant gauge transformation \eqs{A_{\mu}\rightarrow \wt\Omega\,A_{\mu}} with
\eq{\wt\Omega\,\wt\Omega_1\wt\Omega^{-1}=\textrm{Id},\quad \Omega\brc{x}\equiv \wt\Omega\,\wt\Omega_2\brc{x}\wt\Omega^{-1}}
we arrive at
\eqlb{eq:bcA}{\left\{
\begin{array}{l}
A_1\brc{L,t}=\,A_1\brc{0,t},\\
A_1\brc{x,T}=\Omega\brc{x} A_1\brc{x,0},
\end{array}\right.
}
and the consistency condition for $\Omega$ is
\eq{\Omega\brc{0}=\Omega\brc{L} z, \quad z\in \mathbb{Z}_2.}
Now we should be more accurate with the definition of the holonomy around the boundary:
\begin{figure}[H]
  \begin{center}
  \includegraphics[width=4.5cm]{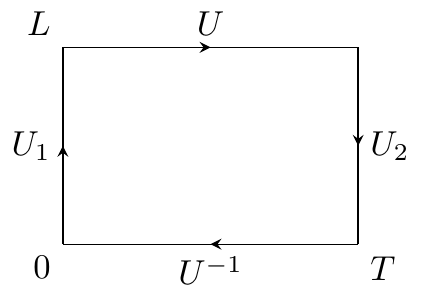}
  \caption{Holonomy around the boundary of the plaquette}
  \end{center}
\end{figure}
For each $U_i$ we have
\eq{U_1=\textrm{Pexp}\brc{\int_{0}^{L}A_1\brc{x,0} \rmd x},}
\eq{U=\textrm{Pexp}\brc{\int_{0}^{T}A_1\brc{L,t} \rmd t},}
\eq{U_2=\textrm{Pexp}\brc{\int_{0}^{1}A_1\brc{x,T} \frac{\rmd x}{\rmd \sigma_2}
\rmd \sigma_2}\quad \textrm{with}\quad x\brc{\sigma_2=0}=L, \,  x\brc{\sigma_2=1}=0.}
Since we have two kinds of vector potentials defined by the boundary conditions with
\eqs{\Omega_0\brc{0}=\Omega_0\brc{L}} and \eqs{\Omega_1\brc{0}=\Omega_1\brc{L}z_1},
\eqs{z_1\neq\textrm{Id}}, the total partition function can be represented as the following sum:
\eq{Z=\frac12\brc{Z_0+Z_1},}
where the factor $1/2$ comes from the normalization of the Haar measure to give volume one. Now, $Z_0$ corresponds to the periodic boundary conditions as in the case of pure $SU\brc{2}$ and we already know the answer:
\eq{Z_0=\sum_{m=0}^{\infty}\rme^{-a\, m\brc{m+2}/4}.}
To compute $Z_1$ we use the boundary conditions to derive
\eq{U_2=\Omega_1\brc{0}\textrm{Pexp}\brc{\int_{0}^{1}A_1\brc{x,0} \frac{\rmd x}{\rmd \sigma_2}
\rmd \sigma_2} \Omega_1^{-1}\brc{L}= \Omega_1\brc{0}U_1^{-1}\Omega_1^{-1}\brc{L}.}
Then the partition function for the cylinder is
\eq{Z_1\brc{a,U_1,U_2}=\sum_{R}\rme^{-a C_2\brc{R}} \chi_{R}\brc{
U_1} \chi_{R}\brc{U_2}.}
Applying the gluing procedure and integrating over $U_1$ we arrive at
\eq{Z_1=\sum_{R}\rme^{-a C_2\brc{R}} \int \rmd U_1 \,\chi_{R}\brc{U_1} \chi_{R}\brc{z_1 U_1^{-1}}
=\sum_{R}\rme^{-a C_2\brc{R}}\dfrac{\chi_R\brc{z_1}}{\textrm{dim} \,R}.}
Using the Weyl character formula for the $SU\brc{2}$ case
\eq{\chi_R\brc{\brc{
\begin{matrix}
\rme^{\imath \phi}& 0\\
0 & \rme^{-\imath \phi}
\end{matrix}
}}=\dfrac{\sin\brc{n\,\phi}}{\sin\brc{\phi}}, \quad n=\textrm{dim} \,R,}
we get $\chi_R\brc{z_1}= n \brc{-1}^{n+1}$ and
\eq{Z_1=\sum_{m=0}^{\infty}\brc{-1}^{m}\rme^{-a\, m\brc{m+2}/4}.}
Thus, the answer for the total partition function is
\eq{Z=\sum_{k=0}^{\infty}\rme^{-a\, k\brc{k+1}},}
which coincides with the general answer (\ref{eq:Z_2D}) for the group $SO\brc{3}$.

\subsection{Adding $w_2$ to \eqs{SU(2)/\mathbb{Z}_2}}
Following \cite{Witten'91,Alekseev'15} we are going to use topological approach to the calculation of the partition function with the second Stiefel–Whitney class $w_2$ added to the Lagrangian:
\eqlb{eq:theta_term}{Z^{\textrm{SW}}=\int \mathcal{D} A\,\rme^{\imath S_{YM}+\imath \theta \, w_2},}
where the dependence on $\theta$ is $2\pi$-periodic and different possible values of theta in the \eqs{SU(2)} case are
\eqs{\theta=0,\pi}.
Since $w_2$ only depends on the topological type of the bundle, the path integral splits into two parts,
corresponding to the trivial and nontrivial $SO(3)$-bundles over the torus. The trivial bundle is defined by the
boundary conditions (\ref{eq:bcA}) with \eqs{\Omega_0\brc{0}=\Omega_0\brc{L}} and the value of $w_2$ is $0$.
The nontrivial bundle is defined by the boundary conditions (\ref{eq:bcA}) with \eqs{\Omega_1\brc{0}=\Omega_1\brc{L}z_1}
and the value of $w_2$ is $1$. In this way we get the following answer:
\eq{Z^{\textrm{SW}}\brc{\theta=\pi}=\frac12\brc{Z_0+\rme^{\imath\pi}Z_1}=\sum_{k=0}^{\infty}\rme^{-a\brc{2k+1}\brc{2k+3}/4},}
where SW stands for Stiefel–Whitney.

\section{\eqs{SU\brc{N}/\mathbb{Z}_N} gauge theory}
\label{sec:PSU(N)}
For the group \eqs{SU(N)/\mathbb{Z}_N} we have $N$ non-equivalent periodic boundary conditions (\ref{eq:bcA}) with
\eq{\Omega_k\brc{0}=\Omega_k\brc{L}z_k, \quad z_k\in\mathbb{Z}_N,\quad k=0,1,\dots, N-1.}
Repeating the steps from the previous section we write the partition function as
\eq{Z=\frac1{N}\sum_{k=0}^{N-1}Z_k,}
where
\eq{Z_k=\sum_{R}\rme^{-a C_2\brc{R}}\dfrac{\chi_R\brc{z_k}}{\textrm{dim} \,R}.}

We can add the $\theta$ term to the Lagrangian as in (\ref{eq:theta_term}) with $w_2$ replaced by an invariant
of $PSU\brc{N}$ bundles \eqs{u_2\in H^2\brc{\mathbf{T}^2;\,\mathbb{Z}_N}} \cite{KSTZ'18}. Allowing theta to take more values inside the \eqs{\left[0,2\pi \right)} interval and labeling these values by $\kappa$, we get
\eq{\theta_{\kappa}=\frac{2\pi\kappa}{N},\quad \kappa=0,\dots,N-1.}
Thus, each $Z_k$ acquires the factor of $\rme^{\imath\theta_{\kappa} k}$ and the corresponding partition function is
\eq{Z_{\kappa}^{\textrm{SW}}\equiv Z^{\textrm{SW}}\brc{\theta=\theta_{\kappa}} =
\frac1{N}\sum_{k=0}^{N-1}\rme^{2\pi\imath \,\kappa\, k/N}Z_k.}

\subsection{Example: \eqs{SU(3)/\mathbb{Z}_3}}
Irreducible representations of \eqs{SU(3)} can be labeled by the Dynkin coefficients \eqs{\brc{n,m}}. The two fundamental weights of \eqs{SU(3)} are
\eq{\mu^{1}=\brc{\frac12,\frac1{2\sqrt{3}}},\quad \mu^2=\brc{0,\frac1{\sqrt{3}}}.}
This gives for the characters of \eqs{z_k} in the representation \eqs{\brc{n,m}}
\eq{\chi_{\brc{n,m}}\brc{z_k}=\textrm{dim}\,R_{\brc{n,m}} \rme^{2\pi\imath\, k\brc{n+2\,m}/3}, \quad k=0,1,2.}
Since
\eq{\sum_{k=0}^{2}\rme^{2\pi\imath\, k\brc{n+2\,m}/3}=3\,\delta\brc{\bsq{n+2\,m} \textrm{mod}\, 3},\quad
\delta\brc{n}\equiv \delta_{n,0}}
and
\eq{C_2\brc{R_{\brc{n,m}}}=\brc{n^2+m^2+n\,m+3\,n+3\,m}/3,}
we derive for the partition function
\eq{Z=\sum_{n,m=0}^{\infty}\rme^{-a\brc{n^2+m^2+n\,m+3\,n+3\,m}/3}\delta\brc{\bsq{n+2\,m} \textrm{mod}\, 3},}
where due to the Kronecker delta function the only non-zero terms are those that have
\eqs{n+2\,m \equiv 0\, \brc{\textrm{mod}\, 3}}.

Adding $u_2$ with \eqs{\theta=\theta_{\kappa}} changes the argument of the delta function by $\kappa$ and we get
\eq{Z_{\kappa}^{\textrm{SW}}=\sum_{n,m=0}^{\infty}\rme^{-a\brc{n^2+m^2+n\,m+3\,n+3\,m}/3}
\delta\brc{\bsq{\kappa+n+2\,m} \textrm{mod}\, 3}.}

\subsection{General case: \eqs{SU(N)/\mathbb{Z}_N}}
Labeling representations of \eqs{SU(N)} by the Dynkin coefficients \eqs{\brc{q_1,\dots,q_{N-1}}\equiv\bld{q}} and using the fundamental weights, we derive for the characters of $z_k$ in the representation \eqs{\brc{q_1,\dots,q_{N-1}}}:
\eq{\chi_{\bld{q}}\brc{z_k}=\textrm{dim}\,R_{\bld{q}} \,\rme^{2\pi\imath\, k\brc{q_1+2\,q_2 +\dots+\brc{N-1}q_{N-1}}/N}, \quad k=0,1,\dots,N-1.}
Then with the help of the simple identity
\eq{\sum_{k=0}^{N-1}\rme^{2\pi\imath\, k\, n /N}=N\,\delta\brc{n\,\textrm{mod}\, N},}
we get for the partition function
\eqlb{eq:Z_N}{Z=\sum_{q_1,\dots,q_{N-1}=0}^{\infty}\rme^{-a\, C_2\brc{R_{\bld{q}}}}\delta\brc{\bsq{\sum_{j=1}^{N-1}j\,q_j} \textrm{mod}\, N},}
where the only non-zero terms are those that have
\eqs{\sum_{j=1}^{N-1}j\,q_j \equiv 0\, \brc{\textrm{mod}\, N}}.
The eigenvalues of the quadratic Casimir operator in (\ref{eq:Z_N}) are given by \cite{FMS'97}
\eq{C_2\brc{R_{\bld{q}}}=\sum_{j,k=1}^{N-1}q_j \brc{q_k +2} G^{k j},}
where \eqs{G^{i j}} is the inverse of the symmetrized Cartan matrix \eqs{G_{i j}} \cite{FS'97}:
\eq{G_{i j}\equiv\frac{8\brc{\alpha_i, \alpha_j}}{\brc{\alpha_i, \alpha_i}\brc{\alpha_j, \alpha_j}}}
and we are using the normalization,  which provides the Killing metric of the form \eqs{g_{ab}=\frac12 \delta_{ab}} and \eqs{\brc{\alpha_i, \alpha_i}=2}.

Adding the usual $\theta$ term with \eqs{\theta=\theta_{\kappa}}, we obtain
\eqlb{Z_SW_N}{Z_{\kappa}^{\textrm{SW}}=\sum_{q_1,\dots,q_{N-1}=0}^{\infty}\rme^{-a\, C_2\brc{R_{\bld{q}}}}\delta\brc{\bsq{\kappa+\sum_{j=1}^{N-1}j\,q_j} \textrm{mod}\, N},
\quad \kappa=1,\dots,N-1.}

\section{Looking for two vacua in \eqs{SU\brc{N}/\Gamma}, \eqs{\Gamma \subset \mathbb{Z}_N} gauge theory}
\label{sec:2vacua}
In this section we will consider the more general case, when the factor group is taken with respect to the subgroup \eqs{\Gamma} of the center of \eqs{SU\brc{N}}. By now we went through several derivations of the partition functions and it is clear what is the generalization of (\ref{Z_SW_N}) for \eqs{\Gamma\neq\mathbb{Z}_N}. If the order of \eqs{\Gamma} is $n$, then we have $n$ non-equivalent periodic boundary conditions (\ref{eq:bcA}) and the corresponding partition function is
\eq{Z_{\kappa}^{\textrm{SW}}=\sum_{q_1,\dots,q_{N-1}=0}^{\infty}\rme^{-a\, C_2\brc{R_{\bld{q}}}}\delta\brc{\bsq{\kappa+\sum_{j=1}^{N-1}j\,q_j} \textrm{mod}\, n},
\quad \kappa=1,\dots,n-1.}

\subsection{\eqs{N=4}}
We start with the explicit answer for the case of \eqs{SU(4)/\mathbb{Z}_4}:
\eq{\left.Z_{\kappa}^{\textrm{SW}}\right|_{\Gamma=\mathbb{Z}_4}=\sum_{q_1,q_2,q_3=0}^{\infty}\rme^{-a\, C_2\brc{R_{\brc{q_1,q_2,q_3}}}}\delta\brc{\bsq{\kappa+\sum_{j=1}^{3}j\,q_j} \textrm{mod}\, 4},
\quad \kappa=1,\dots,3,}
where
\eqlb{eq:eN_4}{C_2\brc{R_{\brc{q_1,q_2,q_3}}}= \frac{1}{8} \left(3\,q_1^2+4\,q_2^2+3\,q_3^2+4\,q_1 q_2+2\,q_1 q_3+ 4\,q_2 q_3 + 12\,q_1+16\,q_2 +12\,q_3\right).}
As it can be checked directly, there is no such value of $\kappa$ that  would produce two vacua.
However, we can also consider the case of \eqs{SU(4)/\mathbb{Z}_2} with \eqs{\kappa=1} and the following partition function:
\eq{\left.Z_{1}^{\textrm{SW}}\right|_{\Gamma=\mathbb{Z}_2}=\sum_{q_1,q_2,q_3=0}^{\infty}\rme^{-a\, C_2\brc{R_{\brc{q_1,q_2,q_3}}}}\delta\brc{\bsq{1+\sum_{j=1}^{3}j\,q_j} \textrm{mod}\, 2}.}
In this case the two vacua contributions are given by \eqs{\bld{q}=\brc{1,0,0}} and \eqs{\bld{q}=\brc{0,0,1}}.

\subsection{General case of \eqs{SU(2 N)/\mathbb{Z}_2} with \eqs{N>1}}
It is easy to show, that the first non-trivial example of $2$d theory with two vacua discussed earlier is just one of the infinite series of \eqs{SU(2 N)/\mathbb{Z}_2} theories with \eqs{N>1}. We again consider the partition function \eqs{\left.Z_{\kappa}^{\textrm{SW}}\right|_{\Gamma=\mathbb{Z}_2}} with \eqs{\kappa=1} or, equivalently, \eqs{\theta=\pi}:
\eq{\left.Z_{1}^{\textrm{SW}}\right|_{\Gamma=\mathbb{Z}_2}=\sum_{q_1,\dots,q_{2N-1}=0}^{\infty}\rme^{-a\, C_2\brc{R_{\bld{q}}}}\delta\brc{\bsq{1+\sum_{j=1}^{2N-1}j\,q_j} \textrm{mod}\, 2}.}
To show that these theories have two vacua we need the following facts about the inverse Cartan matrix
\eqs{G^{i j}}. The first fact is that all elements of this matrix are strictly positive:
\eq{\forall i,j: G^{i j}>0.}
Second, the diagonal elements \eqs{G^{i i}} are given by \cite{Slansky'81,FS'97}:
\eq{G^{i i}=\frac{i\brc{2N-i}}{4N}, \quad i\leq 2N-1.}
And finally, the following relations hold:
\eq{\forall j=2,\dots,N:\quad \sum_{i=1}^{2N-1}G^{i 1}<\sum_{i=1}^{2N-1}G^{i j},}
\eq{\forall j=1,\dots,N-1:\quad \sum_{i=1}^{2N-1}G^{i, N-j}=\sum_{i=1}^{2N-1}G^{i, N+j}.}
Thus, the two vacua contributions are coming from
\eq{\bld{q}=\brc{1,0,\dots,0}\quad \textrm{and}\quad \bld{q}=\brc{0,\dots,0,1}.}
Notice that due to the superselection rules
\eq{\langle R_2 | R_1 \rangle =\int\rmd U \, \chi_{R_1}\brc{U} \chi_{R_2}\brc{U^{-1}}=\delta_{R_1,R_2}}
we indeed have two different ground states that indicate spontaneous symmetry breaking.

\subsection{\eqs{\Gamma\neq\mathbb{Z}_2}}
By studying some particular examples with low enough values of $N$, one can check that
the following theories with \eqs{\Gamma\neq\mathbb{Z}_2} have two vacuum states:
\eqs{SU(8)/\mathbb{Z}_4} with \eqs{\kappa=2}, \eqs{SU(12)/\mathbb{Z}_4} with \eqs{\kappa=2},
\eqs{SU(12)/\mathbb{Z}_6} with \eqs{\kappa=3}, \eqs{SU(16)/\mathbb{Z}_4} with \eqs{\kappa=2},
\eqs{SU(16)/\mathbb{Z}_8} with \eqs{\kappa=4}. Basically, any \eqs{SU(2N\,m)/\mathbb{Z}_{2m}}
theory with \eqs{\kappa=m} and \eqs{N>1}, \eqs{m>1} is a candidate for having two-fold degenerate vacuum.
The only obstacle to making this statement true in general, is that in principle for higher values of $N$ and $m$ there could be states with energies lower than the energy of the following two states:
\eq{q_m=1,\,\forall j\neq m:\, q_j=0 \quad \textrm{and}\quad q_{\brc{2N-1}m}=1,\,\forall j\neq \brc{2N-1}m:\,
q_j=0.}
However, explicit computations for a number of different values of $N$ and $m$ suggest that the above states are always the lowest energy states of the theory with \eqs{\kappa=m}. If we assume that there are states with even lower energies, then they will also come in pairs. This allows us to conclude that the vacuum of the \eqs{SU(2N\,m)/\mathbb{Z}_{2m}} theory with \eqs{\kappa=m} is at least two-fold degenerate. Moreover, the lack of discrete symmetries with order higher than $2$ hints that the two-fold degeneracy is the only option.

\section{Spontaneous symmetry breaking in \eqs{SU(2 N)/\mathbb{Z}_2} theories with \eqs{\theta=\pi}, \eqs{N>1}}
\label{sec:SSB}
If we look at the Dynkin coefficients corresponding to the two vacuum states, we see that these states are given
by the fundamental and antifundamental representations of \eqs{SU(2 N)}. Hence the question is what transformation brings us from one representation to its complex conjugate. Since the wave functions in the propagator
(\ref{eq:propcyl}) are given by
\eq{\langle U | R \rangle =\chi_R\brc{U}=\Tr_R\brc{U},\quad U=\textrm{Pexp}\brc{\int_{0}^{L}A_1\brc{x,t} \rmd x},}
the transformation \eqs{A_1\rightarrow - A_1^T} yields \eqs{U\rightarrow \brc{U^{-1}}^T=U^{*}} and
\eqs{\chi_R\brc{U}\rightarrow\chi_{R^*}\brc{U}}. From the Gauss Law constraint one could derive the following
transformation rules for the \eqs{\textbf{C}}, \eqs{\textbf{P}} and \eqs{\textbf{T}} operators:
\eq{\textbf{C}:\quad e\rightarrow - e,\quad A_1\rightarrow -A_1^T,}
\eq{\textbf{P}:\quad x\rightarrow - x,\quad A_1\rightarrow -A_1,}
\eq{\textbf{T}:\quad t\rightarrow - t,\quad A_1\rightarrow -A_1^T.}
Thus, the \eqs{\textbf{C}}-symmetry (as well as \eqs{\textbf{T}}) is spontaneously broken and the overall \eqs{\textbf{CPT}}-symmetry is conserved, since both \eqs{\textbf{CT}} and \eqs{\textbf{P}} act trivially on the wave functions \eqs{\chi_R\brc{U}}.

Spontaneously broken \eqs{\textbf{C}}- and \eqs{\textbf{T}}-symmetries
lead to the domain wall between the two vacuum states \eqs{\chi_F\brc{U}} and \eqs{\chi_{\bar{F}}\brc{U}}.
In the theories under consideration there is a discrete one-form symmetry $\mathbb{Z}_N$, generated by a local unitary operator \eqs{\mathcal{O}} \cite{GKSW'15,KC'71}. This local operator picks up a phase when crossing the domain wall. To figure out the phase, we consider the $\mathbb{Z}_{2N}$ subgroup before factoring out $\mathbb{Z}_2$. As before, the corresponding characters are given by
\eq{\chi_{\bld{q}}\brc{z_k}=\textrm{dim}\,R_{\bld{q}} \rme^{2\pi\imath\, k\brc{q_1+2\,q_2 +\dots+\brc{2N-1}q_{2N-1}}/\brc{2N}}, \quad k=0,1,\dots,2N-1.}
After factoring out $\mathbb{Z}_2$ the generator of $\mathbb{Z}_N$ corresponds to \eqs{z_1} and its action on the wave functions is simply
\eq{\mathcal{O}| F \rangle=\rme^{\pi\imath/N}| F \rangle,\quad \mathcal{O}| \bar{F} \rangle= \rme^{-\pi\imath/N}|\bar{F}\rangle,}
where \eqs{\mathcal{O}^N=1} due to the fact that $\brc{-1}\in\mathbb{Z}_2$ in fundamental and antifundamental representations.
Here we also assume that adding second Stiefel–Whitney class only affects the $\mathbb{Z}_2$-charges of the states and $\mathbb{Z}_N$-charges remain the same.
In this way the relation between the two expectation values reads
\eq{\langle\mathcal{O}\rangle_{F}=\rme^{2\pi\imath/N}\langle\mathcal{O}\rangle_{\bar{F}},}
which fixes the phase factor picked by \eqs{\mathcal{O}} upon crossing the domain wall to be \eqs{\rme^{2\pi\imath/N}}.
Then the action of charge conjugation on \eqs{\mathcal{O}} can be inferred from
\eq{\textbf{C}\,\mathcal{O}\,\textbf{C}^{-1}| F \rangle=\rme^{-\pi\imath/N}| F \rangle,\quad
\textbf{C}\,\mathcal{O}\,\textbf{C}^{-1}| \bar{F} \rangle=\rme^{\pi\imath/N}| \bar{F} \rangle.}
The latter implies
\eq{\textbf{C}\,\mathcal{O}\,\textbf{C}^{-1}=\mathcal{O}^{-1}.}

\section{Conclusion}
In this note we described a new mechanism of spontaneous symmetry breaking in pure two-dimensional Yang-Mills theories. Using the well-developed methods for computing the $2$d partition functions on compact manifolds \cite{Migdal'75,Witten'91,CMR'95} and G.~'t~Hooft's idea of twisted boundary conditions \cite{Hooft'79},
we analyzed the wide range of systems and presented the corresponding partition functions. We observed that there is no spontaneous symmetry breaking in \eqs{PSU\brc{N}} theories for any $N$, which led us to consider more general case of
\eqs{SU\brc{N}/\Gamma} theories with \eqs{\Gamma} being the subgroup of the center of \eqs{SU\brc{N}}. Within this class of systems we found many examples of theories with degenerate vacuum state and in the particular case of \eqs{SU(2 N)/\mathbb{Z}_2}, \eqs{N>1} we proved that the spontaneous symmetry breaking occurs. Additionally, we argued that the corresponding order parameter is given by the generator \eqs{\mathcal{O}} of the $\mathbb{Z}_N$ one-form symmetry.
There are still a number of questions left to answer. In particular, it will be interesting to prove that the same mechanism of spontaneous symmetry breaking takes place in \eqs{SU(2N\,m)/\mathbb{Z}_{2m}} theories. As we mentioned in the
introduction, the spontaneous symmetry breaking could also imply the existence of the mixed anomaly, which may be a topic of a separate study.

Another interesting question is how to perturb the theories under consideration so that the vacuum is no longer degenerate.
As we discussed earlier, the two vacuum states are given by the fundamental and antifundamental representations. The reason being that the energy of the states is proportional to the eigenvalue of the quadratic Casimir operator, which does not distinguish between any given representation and its complex conjugate. However, if we find a way to modify the theory such that the Hamiltonian will include higher order Casimir operators, we will lift the degeneracy. Indeed, such modifications exist and were discussed in \cite{Witten'91,CMR'95}. Below we will briefly repeat the arguments from \cite{Witten'91,CMR'95} and show how to perturb the Hamiltonian by the cubic Casimir operator.

In $2$d it is possible to define the Lie algebra valued scalar $f=*F$.
Thus, we can add an irrelevant operator $\Tr\brc{f^k}$ with any $k>2$ as a perturbation to the original theory. This perturbation will affect the Hamiltonian by introducing new Casimirs of order less than or equal to $k$ \cite{CMR'95}. Since our goal is to distinguish between the fundamental and antifundamental representations of $SU(N)$, it is enough to consider $k=3$. Then the Hamiltonian is a linear combination of the quadratic and cubic Casimirs (in the representation basis). For example, consider the case of the \eqs{SU(4)/\mathbb{Z}_2} theory. Without the perturbation the energy is proportional to $C_2\brc{R_{\bld{q}}}$ (\ref{eq:eN_4}) and symmetric under the permutation of $q_1$ and $q_3$. In the perturbed system the energy acquires non-zero contributions proportional to
\eqn{C_3\brc{R_{\bld{q}}}=\frac{3}{16} \brc{q_1-q_3} \brc{q_1+q_3+2} \brc{q_1+2\, q_2+q_3+4},}
which is clearly antisymmetric in $q_1$ and $q_3$. Hence, the energies of the two original vacuum states with \eqs{\bld{q}=\brc{1,0,0}} and \eqs{\bld{q}=\brc{0,0,1}} will get different corrections and the resulting system will have
a single vacuum state.
So far we described one possible approach to lifting the degeneracy in the \eqs{SU(2 N)/\mathbb{Z}_2} theories with \eqs{\theta=\pi} and \eqs{N>1}. The existence of any other approaches  and detailed calculations for the case of general $N$ are left for the future work.

\section*{Acknowledgements}
The author would like to thank Zohar Komargodski for initiating the work and many valuable discussions and Konstantinos Roumpedakis for useful remarks and discussions. The work was supported by the "BASIS" Foundation grant 18-1-1-50-3 and in part by RFBR grants 18-02-01081-A, 18-31-20046-mol\_a\_ved, 18-51-05015-Arm\_a, 19-51-18006-Bolg\_a.

\bibliographystyle{unsrt}
\bibliography{references}

\begin{thebibliography}{10}

\bibitem{LS'71}
J.~Lowenstein and J.~Swieca.
\newblock {Quantum electrodynamics in two dimensions}.
\newblock {\em Annals of Physics}, 68(1):172--195, 1971.

\bibitem{Hooft'74}
G.~'t Hooft.
\newblock {A planar diagram theory for strong interactions}.
\newblock {\em Nuclear Physics B}, 72(3):461--473, 1974.

\bibitem{Coleman'76}
S.~Coleman.
\newblock {More about the massive Schwinger model}.
\newblock {\em Annals of Physics}, 101(1):239--267, 1976.

\bibitem{CDG'76}
C.~G. Callan, R.~F. Dashen, and D.~J. Gross.
\newblock {The structure of the gauge theory vacuum}.
\newblock {\em Physics Letters B}, 63(3):334--340, 1976.

\bibitem{Polyakov'77}
A.~M. Polyakov.
\newblock {Quark confinement and topology of gauge theories}.
\newblock {\em Nuclear Physics B}, 120(3):429--458, 1977.

\bibitem{Witten'79}
E.~Witten.
\newblock {Theta-vacua in two-dimensional quantum chromodynamics}.
\newblock {\em Il Nuovo Cimento A}, 51(3):325--338, 1979.

\bibitem{KS'14}
Anton Kapustin and Nathan Seiberg.
\newblock {Coupling a QFT to a TQFT and duality}.
\newblock {\em J. High Energ. Phys.}, 2014(4):1, 2014.
\newblock \href{https://arxiv.org/abs/1401.0740}{arXiv:1401.0740 [hep-th]}.

\bibitem{GKSW'15}
Davide Gaiotto, Anton Kapustin, Nathan Seiberg, and Brian Willett.
\newblock {Generalized Global Symmetries}.
\newblock {\em J. High Energ. Phys.}, 2015(2), 2015.
\newblock \href{https://arxiv.org/abs/1412.5148}{arXiv:1412.5148 [hep-th]}.

\bibitem{Hooft'80}
G.~'t Hooft.
\newblock {Naturalness, Chiral Symmetry, and Spontaneous Chiral Symmetry
  Breaking}.
\newblock {\em NATO Adv. Study Inst. Ser. B Phys.}, 59:135, 1980.

\bibitem{Redlich1'84}
A.~N. Redlich.
\newblock {Parity violation and gauge noninvariance of the effective gauge
  field action in three dimensions}.
\newblock {\em Phys. Rev. D}, 29(10):2366, 1984.

\bibitem{Redlich2'84}
A.~N. Redlich.
\newblock {Gauge Noninvariance and Parity Nonconservation of Three-Dimensional
  Fermions}.
\newblock {\em Phys. Rev. Lett.}, 52(1):18, 1984.

\bibitem{KT1'14}
Anton Kapustin and Ryan Thorngren.
\newblock {Anomalous Discrete Symmetries in Three Dimensions and Group
  Cohomology}.
\newblock {\em Phys. Rev. Lett.}, 112(23):231602, 2014.
\newblock \href{https://arxiv.org/abs/1403.0617}{arXiv:1403.0617 [hep-th]}.

\bibitem{KT2'14}
Anton Kapustin and Ryan Thorngren.
\newblock {Anomalies of discrete symmetries in various dimensions and group
  cohomology}.
\newblock 2014.
\newblock \href{https://arxiv.org/abs/1404.3230}{arXiv:1404.3230 [hep-th]}.

\bibitem{SW'16}
Nathan Seiberg and Edward Witten.
\newblock {Gapped boundary phases of topological insulators via weak coupling}.
\newblock {\em Progress of Theoretical and Experimental Physics}, 2016(12),
  2016.
\newblock \href{https://arxiv.org/abs/1602.04251}{arXiv:1602.04251
  [cond-mat.str-el]}.

\bibitem{BHS'17}
Francesco Benini, Po-Shen Hsin, and Nathan Seiberg.
\newblock {Comments on global symmetries, anomalies, and duality in (2 + 1)d}.
\newblock {\em J. High Energ. Phys.}, 2017(4):135, 2017.
\newblock \href{https://arxiv.org/abs/1702.07035}{arXiv:1702.07035
  [cond-mat.str-el]}.

\bibitem{GKKS'17}
Davide Gaiotto, Anton Kapustin, Zohar Komargodski, and Nathan Seiberg.
\newblock {Theta, time reversal and temperature}.
\newblock {\em J. High Energ. Phys.}, 2017(5):91, 2017.
\newblock \href{https://arxiv.org/abs/1703.00501}{arXiv:1703.00501 [hep-th]}.

\bibitem{TMS'17}
Y.~Tanizaki, T.~Misumi, and N.~Sakai.
\newblock {Circle compactification and 't Hooft anomaly}.
\newblock 2017(12):56, 2017.
\newblock \href{https://arxiv.org/abs/1710.08923}{arXiv:1710.08923 [hep-th]}.

\bibitem{KSTZ'18}
Z.~Komargodski, A.~Sharon, R.~Thorngren, and X.~Zhou.
\newblock {Comments on Abelian Higgs Models and Persistent Order}.
\newblock {\em SciPost Phys.}, 6(1):3, 2019.
\newblock \href{https://arxiv.org/abs/1705.04786}{arXiv:1705.04786 [hep-th]}.

\bibitem{KSU'18}
Z.~Komargodski, T.~Sulejmanpasic, and M.~Unsal.
\newblock {Walls, anomalies, and deconfinement in quantum antiferromagnets}.
\newblock {\em Phys. Rev. B}, 97(5):054418, 2018.
\newblock \href{https://arxiv.org/abs/1706.05731}{arXiv:1706.05731
  [cond-mat.str-el]}.

\bibitem{TKMS'18}
Y.~Tanizaki, Y.~Kikuchi, T.~Misumi, and N.~Sakai.
\newblock {Anomaly matching for the phase diagram of massless
  ${\mathbb{Z}}_{N}$-QCD}.
\newblock {\em Phys. Rev. D}, 97(5):054012, 2018.
\newblock \href{https://arxiv.org/abs/1711.10487}{arXiv:1711.10487 [hep-th]}.

\bibitem{MTU'19}
T.~Misumi, Y.~Tanizaki, and M.~Unsal.
\newblock {Fractional $\theta$ angle, 't Hooft anomaly, and quantum instantons
  in charge-q multi-flavor Schwinger model}.
\newblock {\em Journal of High Energy Physics}, 2019(7):18, 2019.
\newblock \href{https://arxiv.org/abs/1905.05781}{arXiv:1905.05781 [hep-th]}.

\bibitem{CFLS'19}
Clay Cordova, Daniel~S. Freed, Ho~Tat Lam, and Nathan Seiberg.
\newblock {Anomalies in the Space of Coupling Constants and Their Dynamical
  Applications I}.
\newblock 2019.
\newblock \href{https://arxiv.org/abs/1905.09315}{arXiv:1905.09315 [hep-th]}.

\bibitem{CJTU'19}
Aleksey Cherman, Theodore Jacobson, Yuya Tanizaki, and Mithat Unsal.
\newblock {Anomalies, a mod 2 index, and dynamics of 2d adjoint QCD}.
\newblock 2019.
\newblock \href{https://arxiv.org/abs/1908.09858}{arXiv:1908.09858 [hep-th]}.

\bibitem{CO'19}
Clay Cordova and Kantaro Ohmori.
\newblock {Anomaly Obstructions to Symmetry Preserving Gapped Phases}.
\newblock 2019.
\newblock \href{https://arxiv.org/abs/1910.04962}{arXiv:1910.04962 [hep-th]}.

\bibitem{KMS'19}
D.~Kapec, R.~Mahajan, and D.~Stanford.
\newblock {Matrix ensembles with global symmetries and 't Hooft anomalies from
  2d gauge theory}.
\newblock 2019.
\newblock \href{https://arxiv.org/abs/1912.12285}{arXiv:1912.12285}.

\bibitem{WWZ'19}
Zheyan Wan, Juven Wang, and Yunqin Zheng.
\newblock {Quantum 4d Yang-Mills theory and time-reversal symmetric 5d
  higher-gauge topological field theory}.
\newblock {\em Phys. Rev. D}, 100(8):085012, 2019.

\bibitem{WWZ'20}
Zheyan Wan, Juven Wang, and Yunqin Zheng.
\newblock {New Higher Anomalies, SU(N) Yang-Mills Gauge Theory and
  $\mathbb{CP}^{N-1}$ Sigma Model}.
\newblock {\em Annals of Physics}, 414:168074, 2020.

\bibitem{Affleck'91}
Ian Affleck.
\newblock {Nonlinear \ensuremath{\sigma} model at
  \ensuremath{\theta}=\ensuremath{\pi}: Euclidean lattice formulation and
  solid-on-solid models}.
\newblock {\em Phys. Rev. Lett.}, 66(19):2429, 1991.

\bibitem{Sharpe'19}
E.~Sharpe.
\newblock {Undoing decomposition}.
\newblock 2019.
\newblock \href{https://arxiv.org/abs/1911.05080v2}{arXiv:1911.05080}.

\bibitem{PS'05}
T.~Pantev and E.~Sharpe.
\newblock {Notes on gauging noneffective group actions}.
\newblock 2005.
\newblock \href{https://arxiv.org/abs/hep-th/0502027}{arXiv:hep-th/0502027}.

\bibitem{HHPSA'06}
M.~Ando, S.~Hellerman, A.~Henriques, T.~Pantev, and E.~Sharpe.
\newblock {Cluster decomposition, T-duality, and gerby CFT's}.
\newblock {\em Advances in Theoretical and Mathematical Physics}, 11(5):751,
  2007.
\newblock \href{https://arxiv.org/abs/hep-th/0606034}{arXiv:hep-th/0606034}.

\bibitem{Sharpe'14}
E.~Sharpe.
\newblock {Decomposition in diverse dimensions}.
\newblock {\em Phys. Rev. D}, 90(2), 2014.
\newblock \href{https://arxiv.org/abs/1404.3986}{arXiv:1404.3986}.

\bibitem{Migdal'75}
A.~Migdal.
\newblock {Recursion Relations in Gauge Theories}.
\newblock {\em Zh. Eksp. Teor. Fiz.}, 69:810, 1975.

\bibitem{KK'80}
V.A. Kazakov and I.K. Kostov.
\newblock {Non-linear strings in two-dimensional U(\ensuremath{\infty}) gauge
  theory}.
\newblock {\em Nuclear Physics B}, 176(1):199, 1980.

\bibitem{Kazakov'81}
V.A. Kazakov.
\newblock {Wilson loop average for an arbitrary contour in two-dimensional U(N)
  gauge theory}.
\newblock {\em Nuclear Physics B}, 179(2):283, 1981.

\bibitem{AB'83}
M.F. Atiyah and R.~Bott.
\newblock {The Yang-Mills equations over Riemann surfaces}.
\newblock {\em Phil. Trans. R. Soc. A}, 308(1505), 1983.

\bibitem{BZ'85}
K.H. O'Brien and J.-B. Zuber.
\newblock {Strong coupling expansion of large-N QCD and surfaces}.
\newblock {\em Nuclear Physics B}, 253:621, 1985.

\bibitem{Fine'90}
Dana~S. Fine.
\newblock {Quantum Yang-Mills on the two-sphere}.
\newblock {\em Commun. Math. Phys.}, 134(2):273, 1990.

\bibitem{Fine'91}
Dana~S. Fine.
\newblock {Quantum Yang-Mills on a Riemann surface}.
\newblock {\em Commun. Math. Phys.}, 140(2):321, 1991.

\bibitem{Witten'91}
Edward Witten.
\newblock {On quantum gauge theories in two dimensions}.
\newblock {\em Communications in Mathematical Physics}, 141(1):153--209, 1991.

\bibitem{GT'93}
D.~J. Gross and W.~Taylor.
\newblock {Two-dimensional QCD is a string theory}.
\newblock {\em Nuclear Physics B}, 400(1):181, 1993.
\newblock \href{https://arxiv.org/abs/hep-th/9301068}{arXiv:hep-th/9301068}.

\bibitem{Kostov'94}
I.K. Kostov.
\newblock {U(N) gauge theory and lattice strings}.
\newblock {\em Nuclear Physics B}, 415(1):29, 1994.
\newblock \href{https://arxiv.org/abs/hep-th/9308158}{arXiv:hep-th/9308158}.

\bibitem{MP'94}
J.~Minahan and A.~Polychronakos.
\newblock {Classical solutions for two-dimensional QCD on the sphere}.
\newblock {\em Nuclear Physics B}, 422(1):172, 1994.
\newblock \href{https://arxiv.org/abs/hep-th/9309119}{arXiv:hep-th/9309119}.

\bibitem{CMR'97}
S.~Cordes, G.~Moore, and S.~Ramgoolam.
\newblock {Large N 2D Yang-Mills Theory and Topological String Theory}.
\newblock {\em Comm Math Phys}, 185(3):543, 1997.
\newblock \href{https://arxiv.org/abs/hep-th/9402107}{arXiv:hep-th/9402107}.

\bibitem{CMR'95}
Stefan Cordes, Gregory Moore, and Sanjaye Ramgoolam.
\newblock {Lectures on 2D Yang-Mills theory, Equivariant Cohomology and
  Topological Field Theories}.
\newblock {\em Nucl.Phys.Proc.Suppl.}, 41(1-3):184--244, 1995.
\newblock \href{https://arxiv.org/abs/hep-th/9411210}{arXiv:hep-th/9411210}.

\bibitem{Kazakov'94}
Vladimir Kazakov.
\newblock {A String Project in Multicolour QCD}.
\newblock {\em String Theory, Gauge Theory and Quantum Gravity '93}, page~29,
  1994.
\newblock \href{https://arxiv.org/abs/hep-th/9308135}{arXiv:hep-th/9308135}.

\bibitem{BT'91}
Matthias Blau and George Thompson.
\newblock {Quantum Maxwell theory on arbitrary surfaces}.
\newblock 1991.
\newblock (NIKHEF-H/91-08).

\bibitem{BT'92}
Matthias Blau and George Thompson.
\newblock {Quantum Yang-Mills theory on arbitrary surfaces}.
\newblock {\em International Journal of Modern Physics A}, 7(16):3781, 1992.

\bibitem{BT'93}
Matthias Blau and George Thompson.
\newblock {Lectures on 2d Gauge Theories: Topological Aspects and Path Integral
  Techniques}.
\newblock 1993.
\newblock \href{https://arxiv.org/abs/hep-th/9310144}{arXiv:hep-th/9310144}.

\bibitem{Hooft'79}
G.~'t Hooft.
\newblock {A property of electric and magnetic flux in non-abelian gauge
  theories}.
\newblock {\em Nuclear Physics B}, 153:141--160, 1979.

\bibitem{Alekseev'15}
Anton Alekseev, Olga Chekeres, and Pavel Mnev.
\newblock {Wilson surface observables from equivariant cohomology}.
\newblock {\em J. High Energ. Phys.}, 2015(11), 2015.
\newblock \href{https://arxiv.org/abs/1507.06343}{arXiv:1507.06343 [hep-th]}.

\bibitem{FMS'97}
P.~Di~Francesco, P.~Mathieu, and D.~Senechal.
\newblock {\em Conformal Field Theory}.
\newblock Springer-Verlag, 1997.

\bibitem{FS'97}
J.~Fuchs and C.~Schweigert.
\newblock {\em Symmetries, Lie Algebras and Representations}.
\newblock Cambridge University Press, 1997.

\bibitem{Slansky'81}
R.~Slansky.
\newblock {Group theory for unified model building}.
\newblock {\em Physics Reports}, 79(1), 1981.

\bibitem{KC'71}
Leo~P. Kadanoff and Horacio Ceva.
\newblock {Determination of an Operator Algebra for the Two-Dimensional Ising
  Model}.
\newblock {\em Phys. Rev. B}, 13(11):3918, 1971.

\end{thebibliography}

\end{document}